\documentclass[lettersize,journal]{IEEEtran}
\usepackage{amsmath,amsfonts}
\usepackage{algorithmic}
\usepackage{algorithm}
\usepackage{array}
\usepackage[caption=false,font=normalsize,labelfont=sf,textfont=sf]{subfig}
\usepackage{textcomp}
\usepackage{stfloats}
\usepackage{url}
\usepackage{verbatim}
\usepackage{graphicx}
\usepackage{cite}

 \usepackage{booktabs}
 
\begin{document}

\title{Numerical and Theoretical Investigation of Multi-Beam Interference and Cavity Resonance in Top-Emission QLEDs}

\author{Hyuntai Kim, and Seong-Yong Cho
     \thanks{Hyuntai Kim is with Electrical and Electronic Convergence Department, Hongik University, Sejong, 30016, Republic of Korea. (e-mail: hyuntai@hongik.ac.kr).}%
     \thanks{Seong-Yong Cho is with Department of Photonics and Nanoelectronics, BK21 FOUR ERICE-ACE Center, Hanyang University ERICA, Ansan, 15588, Korea. (e-mail: seongyongcho@hanyang.ac.kr)}%
    
}

%\markboth{Journal of Lightwave Technology,~Vol.~XX, No.~X, March~2025}%

%\IEEEpubid{0000--0000/00\$00.00~\copyright~2024 IEEE}

\maketitle

\begin{abstract}
Top-emission quantum dot light-emitting diodes (QLEDs) have been extensively studied due to their potential application in augmented/virtual reality. Particularly, the impact of Fabry-Pérot resonance on top-emission QLEDs has been investigated through both experimental and theoretical studies. Additionally, multi-beam interference effects in QLED emission layers have been explored theoretically. However, previous studies predominantly rely on simplified simulations or governing equations with minor numerical corrections, often resulting in discrepancies between theoretical predictions and experimental results. Notably, a comprehensive investigation of multi-beam interference effects remains insufficient.

This study aims to perform a theoretical analysis of multi-beam interference, substantiated with numerical simulations. Specifically, we examine Fabry-Pérot resonance effects and compare them with interference between upward and downward emission components in QLED layers. The findings are expected to provide insights into designing more efficient QLED architectures.

\end{abstract}


\section{Introduction}
\IEEEPARstart{A}{s} the demand for high value-added displays, such as augmented reality (AR) and virtual reality (VR), continues to grow, top-emitting display structures are receiving increasing attention.\cite{lee2024top}  In a top-emitting display structure, a higher aperture ratio can be achieved as the backplane areas occupied by transistors and capacitors are fully utilized for light emission. In particular, ultrahigh-definition quantum dots (QD)-based displays are being actively studied on silicon (Si) substrates with CMOS (complementary metal-oxide-semiconductor) backplanes instead of conventional glass substrates.\cite{mei2022full, li2023highly, rhee2020recent} High-resolution patterning of the QD emission layer using transfer techniques and optical lithography has been actively studied for the fabrication of pixels in AR/VR displays.\cite{yoo2024highly, lee2023high}  In addition, the use of Si substrates is expected to facilitate the fabrication of ultra-high-definition displays by leveraging high-speed transistors rather than thin-film transistors on glass substrates. \cite{park2023progress, kang2024advances} 

  However, in such top-emitting displays, the luminance is expected to vary dramatically depending on the thickness of the functional layers and transparent electrode. \cite{chen2022ultrahigh,li2023highly, lee2024top} The importance of the thickness of the microcavity, which induces constructive and destructive interference between light reflected from the underlying substrate and light emitted towards the front, cannot be overstated.\cite{zhu2014light, lee2024recent} Moreover, the varying optimal thickness conditions required to maximize light extraction in red, green, and blue pixels present a significant challenge in the research and development of top-emitting devices.\cite{mei2022full, qdh1} In this regard, multi-beam interference effects, which depend on the emission wavelength (red, green, and blue), warrant further investigation. \cite{li2023highly,kang2016generalized} 

  In this study, we investigated the Fabry-Pérot resonance and multi-beam interference characteristics and a QD-based top-emitting structure by varying the thicknesses of the bottom electrode (ITO) and the top ZnO layer while employing a standard device configuration (anode/hole injection layer/hole transport layer/emissive layer/electron transport layer/cathode). \cite{li2024ultrabright, shi2022over} The goal of this research is to maximize the efficiency of top emission by optimizing resonance effects.

Fabry-Pérot resonance has been widely discussed in previous studies \cite{zhang2023taming, chen2020high}, and the effects of multi-beam interference have also been extensively investigated \cite{liu2019high, mahissi2023diaphragms}. However, relatively few studies have systematically aligned analytically derived theoretical results with numerically computed simulation outcomes. In this study, a comprehensive analysis is conducted to reconcile theoretical calculations with numerical simulations. Through this process, the respective influences of Fabry-Pérot resonance and multi-beam interference are examined in detail, providing deeper insights into their roles in optical performance.

\section{Principle and Method}
\subsection{Theoretical Analysis}

\begin{figure}[!htb]
\centering
\includegraphics[width=0.95\linewidth]{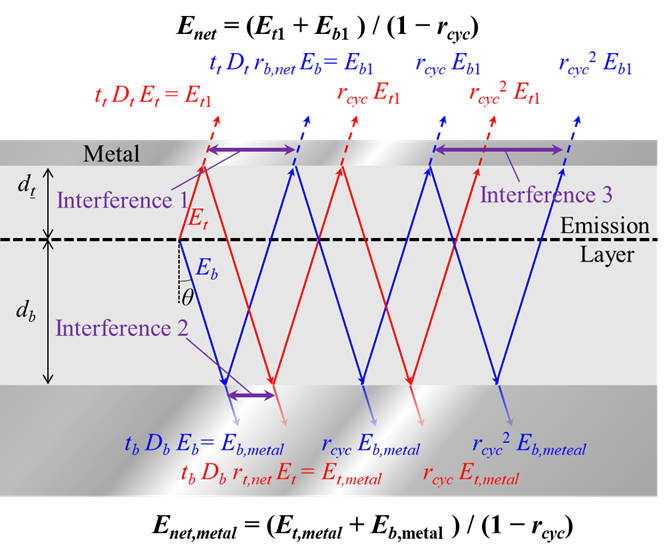}
\caption{Schematic Representation of Light Propagation and Interference in a Top-Emission QLED.}
\label{f1}
\end{figure}

In top-emission QLEDs, light propagates bidirectionally within the emission layer, generating both upward and downward components. Simultaneously, these optical waves undergo multiple reflections at the upper and lower interfaces, leading to complex interference effects. A schematic representation of these interference mechanisms is depicted in Fig.\ref{f1}.

Here, $E_t$ and $E_b$ represent the upward and downward emission components from the emission layer, respectively. $D_t$ and $D_b$ denote the phase shifts introduced by the upper and lower layers, where $D_t = exp(i k d_t)$ and $D_b = exp(i k d_b)$. $k$ is the wavenumber inside the cavity, represented as $k = 2n\pi/\lambda $, where $n$ is the refractive index inside the cavity. The parameters $r_b$, $r_t$, $t_t$ and $t_b$ correspond to the reflection and transmission coefficients of the bottom and top layers, respectively. The first transmitted components are denoted by $E_{t1}$ and $E_{b1}$. $E_{t,metal}$ and $E_{b,metal}$ represents the first transmitted field at the bottom side, before decay. The terms $r_{t,net}$ and $r_{b,net}$ represent the cumulative effects of reflections at the top and bottom layers, while $r_{cyc}$ denotes the effect of a full cavity round trip.

The representation of the top emission component is denoted by a red arrow. It propagates through the top layer, thickness of $d_t$, undergoing a phase change given by $\exp(i k d_t)$, which we define as $D_t$. Assuming the top metal layer is sufficiently thin, both reflection and transmission occur. The transmitted electric field can then be expressed as $t_t D_t E_t$, and the reflected electric field as $r_t D_t E_t$. 

This electric field propagates further to the bottom layer, acquiring an additional phase shift of $D_t D_b$. Upon encountering the bottom layer, it reflects with a coefficient $r_b$ and re-propagates to the top layer, incurring another phase shift of $D_t D_b$. After completing a round trip, the electric field is characterized by a cumulative phase and reflection coefficient product of $r_t r_b D_t^2 D_b^2$. We designate this net change in the electric field due to the cyclic path as $r_{cyc}$. 

The bottom emission, represented by the blue arrow, exhibits similar phenomena. It propagates to the bottom layer, acquiring phase shift of $D_b$. At the bottom layer, we assume that the metallic layer is thick enough that there are no transmission. However, we note that electric field exists at the surface, and it decays while propagating. The electric field at the boundary of the metal region can be also calculated, and the value is $t_b D_b E_b$. We express this term as $E_{b,metal}$. 

The final top layer electric field $E_{net}$ transmited topward and the net electric field at the bottom layer surface $E_{net,metal}$ becomes 
\begin{equation}
%E_{net}=(E_{t1} + E_{b1}) (1 + r_{cyc}+ r_{cyc}^2+ r_{cyc}^3+...) = (E_{t1} + E_{b1}) / (1-r_{cyc})
E_{net} = (E_{t1} + E_{b1}) / (1-r_{cyc})
\label{eq:enet}
\end{equation}
\begin{equation}
E_{net,metal}=(E_{b,metal} + E_{t,metal}) / (1-r_{cyc}).
\label{eq:enetm}
\end{equation}
If we assume that the top emission component and bottom emission components are same, (i.e. $E_t = E_b$), $E_{b1} = r_{b,net} E_{t1}$  and $E_{t,metal} = r_{t,net} E_{b,metal}$. The total field becomes $E_{net} = E_{t1} (1 + r_{b,net} )/(1-r_{cyc})$ and $E_{net,metal} = E_{t,metal} (1 + r_{t,net} )/(1-r_{cyc})$.
Here, we can also define $t_{net}$ and $t_{net,metal}$, which is the ratio between incidence $E_t$ and the outward electric field $E_{net}$ and $E_{net,metal}$.

Three types of interference can be identified within the top emission QLED. The first interference occurs between the upward emission component and the downward component reflected from the bottom layer. The second interference arises from the interaction between the downward emission component and the upward component reflected from the top layer. The final interference results from the self-interference of light after completing a full cavity round trip. These interferences are labeled as Interference 1, 2, and 3 in Fig.\ref{f1}, respectively.

The phase matching condition of interference 1 can be expressed mathematically by $2 k d_b \cos(\theta) + \arg(r_b) = 2 m \pi$ and is satisfied when the phase condition $\arg(r_{b,\text{net}}) = 0$ holds. Similarly, in-phase of interference 2 is generally described by $2 k d_t \cos(\theta) + \arg(r_t) = 2 m \pi$ and is valid under the condition $\arg(r_{t,\text{net}}) = 0$.
Finally, phase matching condition of interference 3, also known as Fabry-Pérot resonance condition can be formulated by $2 k (d_b + d_t) \cos(\theta) + \arg(r_b r_t) = 2 m \pi$ and is also satisfied when $\arg(r_{t,\text{net}}) = 0$. These equations define constructive interference when their values are multiples of $2\pi$ and destructive interference when they are $(2m+1)\pi$. 

A significant limitation in previous studies is the assumption of a plane wave or dipole radiation source within the cavity. While external field reflections can be approximated as plane waves, internal plane wave sources lead to continuous field buildup, causing erroneous results such as transmission exceeding unity. This issue is exacerbated under resonance conditions, leading to an unrealistic accumulation of energy.

To resolve this, we introduce the concept of the transmission-absorption ratio. The transmission coefficient through a metallic layer is typically defined as $|t^2| (n_1/n_2)$, but this is unsuitable for metals with significant absorption. Instead, we define the absorption coefficient as $A = 1 - R$ for an infinitely thick metal, and $A = 1 - R - T$ for thin films. By considering the ratio of transmitted field intensity to absorbed energy, we accurately estimate the top-emission efficiency based on energy conservation law.

In this model, the transmission - absorption ratio becomes $TAR_t = |t_t|^2/(1-|r_t|^2-|t_t|^2/n)$ for top layer and $TAR_b = |t_b|^2/(1-|r_b|^2)$ for bottom layer. Defining the transmission of top layer, absorption of top layer and bottom layer as $T_{net}, A_{t}$ and $A_{b}$, the total transmission efficiency becomes

\begin{equation}
  \frac{T_{net}}{T_{net}+A_{t}+A_{b}}=  \frac{|t_{net}|^2/n}{|t_{net}|^2(1/n+TAR_t)+|t_{net,metal}|^2TAR_b}
  \label{eq:effnet}
\end{equation}

where $t_{net} = E_{net}/E_t$, $t_{net,metal} = E_{net,metal}/E_t$. From the given equation, it is evident that the term  $1/(1-r_{cyc})$ of the $t_{net}$ and $t_{net,metal}$ appears in both the numerator and the denominator, leading to its cancellation. As the numerator is proportional to $t_{net}$, it contains $1-r_{b,net}$ term, it shows the interference 1 is the dominant condition of the transmission. It is also notable that $t_{net,metal}$ term is only at the denominator, so to maximize the transmission, interference 2 must be minimized. 

This finding deviates from conventional assumptions, indicating that the effect of Fabry-Pérot resonance is not a dominant factor. In addition, interference 2, which occurs when the downward emission component interacts with the upward component reflected from the top interface at the bottom metal layer, achieves maximum efficiency when it undergoes destructive interference. This aligns with the physical principle that, regardless of the Fabry-Pérot resonance condition, an internal emission source must conserve energy through either metal absorption or top-side transmission, rendering the cavity resonance condition negligible. Furthermore, to minimize optical losses at the bottom interface, the field intensity near the lower metal layer should be minimized. Consequently, interference 2 leading to destructive interference at this interface is physically justified, as it effectively reduces unnecessary field accumulation and associated losses. Interference 1 must be maximized to optimize transmission, which remains consistent with conventional understanding. The phase-matching condition to optimize the top emission QLED proposed in this work is shown as following:

\begin{equation}
    2 k d_b \cos(\theta) + \arg(r_b) = 2 m \pi
    \label{eq:con}
\end{equation}
\begin{equation}
    2 k d_t \cos(\theta) + \arg(r_t) = (2 m + 1 ) \pi
    \label{eq:des}
\end{equation}

These two equations exhibit similarities to those in previous studies.\cite{zhang2023taming, chen2020high} In particular, Eq.~\ref{eq:con} is mathematically equivalent to existing formulations; however, two notable distinctions exist. First, the absence of an equation related to Fabry-Pérot resonance implies that the total cavity length does not play a significant role. Second, Eq.~\ref{eq:des} follows a phase condition of \( (2m+1)\pi \) rather than \( 2m\pi \), which can be attributed to the use of a metallic layer as the reflective medium at the bottom interface. To minimize optical losses at the metal surface, the multi-beam interference effect corresponding to interference 2 must exhibit destructive rather than constructive interference. Naturally, this consideration becomes irrelevant if the bottom layer is an approximately lossless mirror, such as an ideal perfect conductor or a precisely engineered distributed Bragg reflector (DBR) \cite{dbr1,dbrled}.

\subsection{Numerical Simulation Method}
\begin{figure}[htb]
\centering
\includegraphics[width=0.8\linewidth]{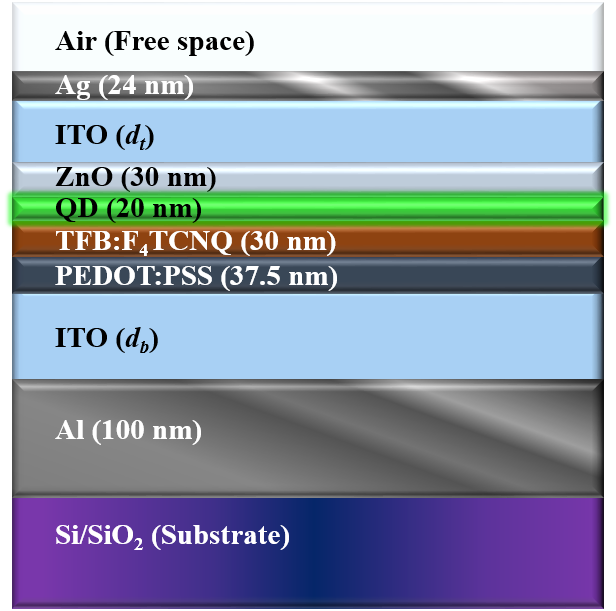}
\caption{Structure of the QLED.}
\label{fig:str}
\end{figure}

The simulation geometry follows conventional top-emission QLED structures, consisting of a 100 nm thick aluminum (Al) reflective electrode, a controllable thickness $d_b$ of ITO layer, a 37.5 nm PEDOT:PSS [Poly(2,3-dihydrothieno-1,4-dioxin)-poly(styrenesulfonate)] layer, a 30 nm TFB (Poly[(9,9-dioctylfluorenyl-2,7-diyl)-co-(4,4'-(N-(4-sec-butylphenyl)diphenylamine)]) layer, a 20 nm QD layer, 30 nm of ZnO layer, another ITO layer with variable thickness $d_t$, and a 24 nm silver (Ag) top electrode. The thickness of each functional layer was fixed, including PEDOT:PSS, TFB:F4TCNQ, QD, and ZnO due to electrical charge transport. Instead, bottom and top ITO thicknesses are expected to affect light outcoupling significantly. The thickness of the top electrode, Ag, was fixed at 24 nm, as this was determined to maintain a high optical transmittance while ensuring a reasonable level of electrical resistance. If the thickness was increased, the optical transmittance would drop sharply, whereas if it were too thin, the resistance would increase, preventing it from functioning effectively as an electrode. The operation wavelength $\lambda$ has been selected to be 520 nm, which is a typical value of green QLED. The structure of the QLED is shown in Fig. \ref{fig:str}, and the optical properties of these layers are detailed in Table \ref{ri}. \cite{npss,nag,nal,shi2022over}
\begin{table}[h]
    \centering
    %\scriptsize
    \begin{tabular}{l c}
        \hline
        Material & Refractive Index \\
        \hline
        ZnO & 1.9639 \\
        TFB & 1.85 \\
        QD & 1.88 \\
        PEDOT:PSS & 1.5201 \\
        Ag & 0.077 + 2.9962i \\
        Al & 0.52153 + 4.9843i \\
        ITO & 1.8933 + 0.0035703i \\
        \hline
    \end{tabular}
    \caption{Refractive indices of materials used in the study.}
    \label{ri}
\end{table}

Finite element method (FEM) via COMSOL Multiphysics was applied to analyze the optical properties of the QLED structure. Mesh sizes were selected as follows: thin layers below 10 nm were meshed with a resolution of $\lambda/100$, while other layers were meshed at $\lambda/60$. Transmission efficiency was computed by integrating the power emitted in the upper direction while accounting for lateral and rearward scattering losses.The emission layer was modeled as a QD layer with a Gaussian-distributed current source, representing the spatial distribution of emitters.

\section{Results and Discussion}
\begin{figure*}[!htb]
\centering
\includegraphics[width=0.8\linewidth]{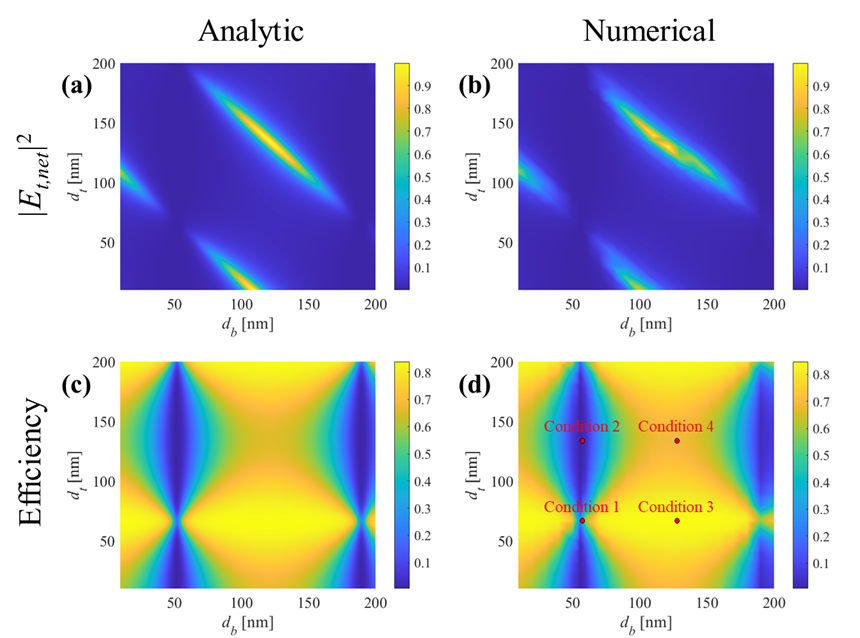}
\caption{(a) Theoretical and (b) numerical results for top emission transmission. (c) Theoretical and (d) numerical results of the top emission efficiency considering energy conservation.}
\label{fig:tnr}
\end{figure*}

Based on the previously established theoretical framework and numerical methodologies, we obtained both analytical and simulation-based results. Figure \ref{fig:tnr} (a) and (b) presents the theoretical and numerical results for electric field transmission intensity, $|E_{net}|^2$ component. Figure \ref{fig:tnr}(c) and (d) shows the total efficiency considering the energy-conserving of the top-emission, described in Eq. \ref{eq:effnet}.

One key observation is that the results of numerical FEM simulations shown in Fig. \ref{fig:tnr}(b) also exhibit field accumulation similar to the theoretical results in Fig. \ref{fig:tnr}(a) when an internal source is introduced within the cavity. 

A more critical observation is that Fig.\ref{fig:tnr}(c) and Fig.\ref{fig:tnr}(d) do not exhibit the characteristic Fabry-Pérot pattern along the diagonal direction, where $d_{t} + d_{b}$ remains constant. Instead, the emission efficiency is primarily influenced by $d_{b}$ and $d_{t}$ independently, a trend consistently observed in both theoretical and numerical results.

The theoretically calculated thicknesses corresponding to the maximum and minimum transmission are as follows. For \( d_t \), destructive interference resulted in a minimum transmission at 52.2 nm, while constructive interference led to a maximum transmission at 120.9 nm. Similarly, for \( d_b \), destructive interference yielded maximum efficiency at 66.3 nm, whereas constructive interference produced the lowest efficiency at 135 nm.  

The simulation results closely align with these theoretical predictions. The minimum and maximum transmission for \( d_t \) were observed at 56 nm and 121 nm, respectively, while for \( d_b \), the minimum and maximum efficiency were found at 67 nm and 136 nm, respectively. This high degree of agreement between theoretical and numerical results validates the theoretical model presented in this study.  

Furthermore, the maximum and minimum efficiency values also exhibited strong consistency. The theoretical model predicted a maximum efficiency of 83.91\% and a minimum efficiency of 0.42\%, while the simulation yielded corresponding values of 84.7\% and 0.52\%, with an error margin of less than 1 percentage point. These findings demonstrate that the theoretical calculations presented in this study are highly consistent with the simulation results, supporting the validity and reliability of the proposed analytical model.

%These findings confirm that, unlike external light sources where Fabry-Pérot effects dominate, internal emission sources result in energy escaping primarily through interference conditions rather than resonance conditions. This highlights the dominant role of multi-beam interference in optimizing top-emission QLED performance.

\begin{figure*}[!htb]
\centering
\includegraphics[width=0.8\linewidth]{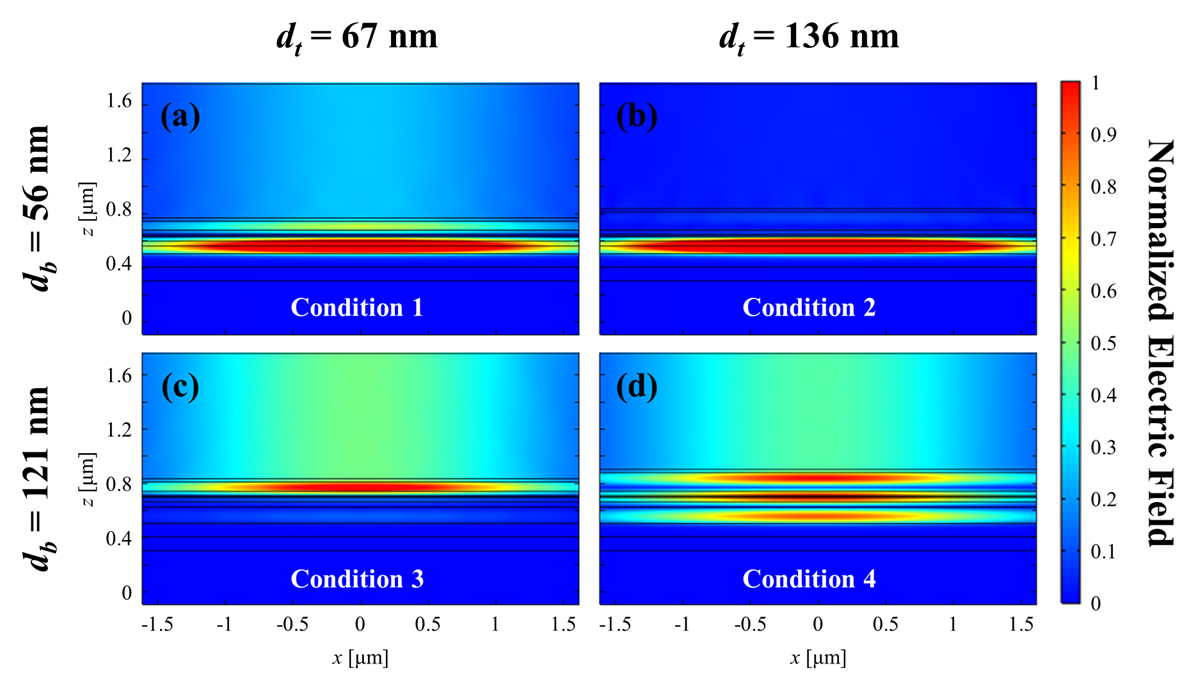}
\caption{The electric field magnitude of critical points.}
\label{fig:fp}
\end{figure*}

To analyze the field distribution, the electric field intensity \( |E|^2 \) at four critical conditions  are presented in Fig.\ref{fig:fp}, where the four points are also depicted in Fig. \ref{fig:tnr}(d) as Conditions 1 to 4. The top ITO thickness, \( d_t \), is set to 67 nm for Figs.\ref{fig:fp}(a) and \ref{fig:fp}(c), where the bottom interference (interference 2) exhibits constructive interference. Conversely, in Figs.\ref{fig:fp}(b) and \ref{fig:fp}(d), \( d_t \) is increased to 136 nm, corresponding to a condition where the bottom interference is minimized. Similarly, the bottom ITO thickness, \( d_b \), is 56 nm for Figs.\ref{fig:fp}(a) and \ref{fig:fp}(b), and 121 nm for Figs.\ref{fig:fp}(c) and \ref{fig:fp}(d). The condition \( d_b = 56 \) nm corresponds to constructive interference at the top interface (interference 1), while \( d_b = 136 \) nm corresponds to destructive interference.

When constructive interference occurs at the top interface, the electric field is predominantly confined near the upper region, whereas destructive interference at this interface results in reduced field intensity in the upper region. A similar trend is observed for the bottom interface: when interference is maximized, the field becomes concentrated within the bottom layer. From the field distribution patterns, it is evident that a higher field concentration near the bottom layer leads to reduced transmission. This is reflected in Fig.\ref{fig:fp}(a), which exhibits lower transmission compared to Fig.\ref{fig:fp}(c), and in Fig.\ref{fig:fp}(b), which shows reduced efficiency compared to Fig.\ref{fig:fp}(d). Conversely, when the field is concentrated in the upper region, the top-emission efficiency improves. As a result, Figs.\ref{fig:fp}(a) and \ref{fig:fp}(c) demonstrate higher efficiency compared to Figs.\ref{fig:fp}(b) and \ref{fig:fp}(d), respectively.

%To observe the field pattern, the electric field magnitude $|E|^2$ of the four critical points are shown in Fig. 4. The top ITO thickness $d_t$ is 67 nm for Figs. 4(a) and 4(b), where the bottom interference (intereference 2) is constructive. The $d_t$ is 136 nm for Figs. 4(c) and (d), where the bottom interference is minimized. Similar to the bottom ITO thickness, $d_b$ is 56 nm for Figs. 4(a) and (c), and 121 nm for Figs. 4(b) and (d). The condition $d_b$ = 56 nm matches the top interference (interference 1) constructive, and 136 nm destructive. Where the top interference is constructive, one can observe the field is confined at the upper side, and vice versa. Similar to the bottom interference, when the bottom interference is maximized, field is confined at the bottom layer. From the field pattern, it is clear that if there are more field concentrated at the bottom layer, the transmission becomes lower. Figure 4(a) shows less transmission than Fig. 4(c), and Fig. 4(b) shows lower efficiency compare to Fig. 4(d). When the field is concentrated at the upper region, the top emission efficiency becomes higher. Figures 4(a) and 4(c) shows higher efficiency compare to Figs. 4(b) and 4(d), respectively.

In this study, although the thickness of the top ITO layer was varied, the same effect can be achieved by modifying layers other than the ITO layer. In other words, the length above the emission layer and the length below it influence the interference at the top and bottom interfaces, respectively. In other LED devices or optical cavities, it is also possible to control the thickness of a lossless medium other than ITO to achieve similar effects.

\begin{figure}[!htb]
\centering
\includegraphics[width=0.95\linewidth]{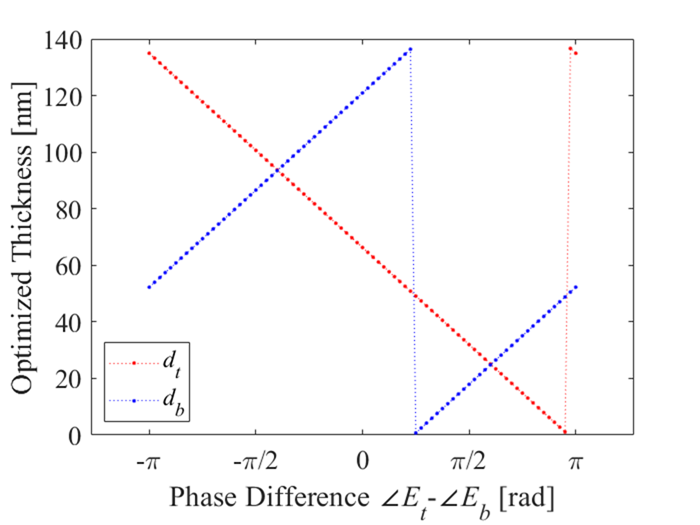}
\caption{Optimized values as a function of the phase difference between the upward and downward emitted light. }
\label{fdp}
\end{figure}

Another aspect to consider is that the magnitude and phase of \( E_t \) and \( E_b \) may not always be identical. Various factors, such as the anisotropy in the structure of QDs and quantum mechanical energy distributions, can contribute to discrepancies between the upward and downward emission components. These variations may arise due to differences in dipole orientation, non-uniform charge carrier distributions, or structural inhomogeneities within the emission layer. However, despite these potential deviations, the fundamental conclusion remains unchanged: while such factors may lead to differences in the optimal layer thickness for maximizing efficiency, the individual thicknesses of \( d_t \) and \( d_b \) remain the most critical parameters governing the overall interference effects. This indicates that even when emission asymmetry is present, the impact of top and bottom layer thicknesses on device performance follows the same essential trends. Therefore, controlling \( d_t \) and \( d_b \) with high precision is crucial for optimizing the emission characteristics of top-emission QLEDs, regardless of variations in emission symmetry. The optimized value, influenced by the phase difference between the upward and downward emitted light, $\angle E_t - \angle E_b$ , is presented in Fig. \ref{fdp}.

\begin{figure}[!htb]
\centering
\includegraphics[width=0.9\linewidth]{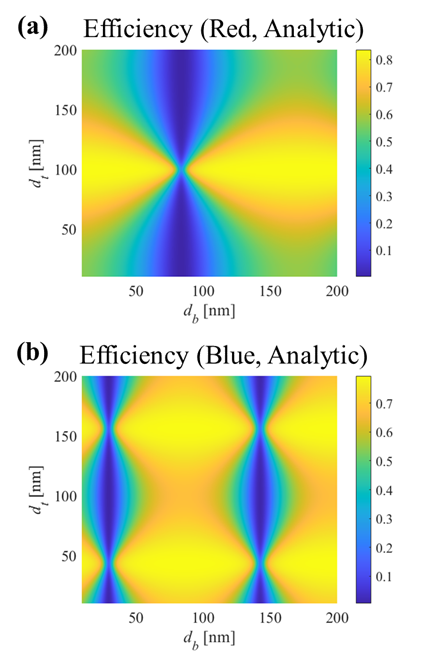}
\caption{Theoretically calculated top emission efficiency for (a) red and (b) blue wavelength.}
\label{fig:rb}
\end{figure}

Theoretical calculations were also conducted for different wavelengths, specifically red (620 nm) and blue (445 nm). The corresponding results are presented in Fig.~\ref{fig:rb}. As illustrated in the figure, the overall interference patterns for these wavelengths exhibit a similar trend to that of green. In the case of red, the longer wavelength led to an increased periodicity, whereas for blue, the periodicity was correspondingly reduced. Additionally, the optimal structural parameters for all three wavelengths, along with the theoretically and numerically computed efficiencies, are summarized in Table~\ref{tab:efficiency}. Note that $Eff_{th}$ and $Eff_{num}$ represent the efficiencies obtained from theoretical calculations and numerical simulations, respectively.

\begin{table}[h]
    \centering
    \begin{tabular}{c|ccc}
        \hline
        Parameter & Red (620 nm) & Green (520 nm) & Blue (445 nm) \\
        \hline
        $d_b$ (nm) & 170.3 & 120.9 & 86.4 \\
        $d_t$ (nm) & 99.6 & 66.3 & 43.3 \\
        $\text{Eff}_{\text{th}}$ (\%) & 83.6 & 83.9 & 79.4 \\
        $\text{Eff}_{\text{nu}}$ (\%) & 84.5 & 84.7 & 80.3 \\
        \hline
    \end{tabular}
    \caption{Optimal thickness values ($d_b$ and $d_t$) and efficiency ($\text{Eff}_{\text{th}}$ and $\text{Eff}_{\text{nu}}$) for red, green, and blue wavelengths.}
    \label{tab:efficiency}
\end{table}

For the red wavelength, the optimal bottom ITO thickness is significantly larger (170.3 nm) compared to other wavelengths. This indicates that further reduction of the bottom layer, along with a decrease in ITO thickness, could result in an optimal condition at a lower thickness. Even if achieving such precise control is challenging, minimizing the ITO thickness alone would still contribute to maintaining a relatively high efficiency.

\section{Conclusion}

In this study, we have conducted a comprehensive theoretical and numerical analysis of the optical interference mechanisms in top-emission QLEDs. Our findings reveal that, contrary to conventional assumptions, Fabry-Pérot resonance does not play a dominant role in determining emission efficiency when the emission source is located within the cavity. Instead, multi-beam interference between upward and downward emission components significantly influences the optical characteristics of top-emission QLEDs.

Through systematic theoretical derivation and numerical simulations, we have demonstrated that conventional Fabry-Pérot resonance effects, which are typically considered in external illumination cases, do not lead to substantial variations in emission efficiency. The absence of a clear diagonal periodicity in the transmission results, as seen in both theoretical and numerical calculations, supports this conclusion. Instead, the emission characteristics are primarily governed by the interference conditions at the upper and lower layers, with the bottom layer favoring in-phase constructive interference and the top layer exhibiting optimal efficiency under out-of-phase conditions.

Additionally, our results highlight the importance of considering the transmission-absorption ratio as a more accurate metric for evaluating emission efficiency. Traditional approaches based on transmission field calculations often fail to account for the loss mechanisms within metallic layers, leading to inconsistencies between theoretical predictions and actual device performance. By introducing a loss-aware approach, we provide a more reliable framework for optimizing top-emission QLED structures.

These insights provide valuable guidelines for designing high-efficiency top-emission QLEDs and other LED-based devices. Specifically, our study suggests that optimizing the thickness of the top and bottom layers should prioritize interference control rather than cavity resonance tuning. Furthermore, our findings indicate that asymmetries in the QD emission profile, such as dipole orientation and quantum mechanical energy states, may influence the optimal design parameters but do not fundamentally alter the primary role of multi-beam interference in determining emission characteristics.

%Future research should focus on experimental validation of these findings, particularly through direct measurements of interference effects and loss mechanisms in practical QLED structures. Additionally, extending this analysis to other optical devices that rely on internal emission sources, such as OLEDs and perovskite LEDs, could provide broader insights into the interplay between resonance and interference in thin-film photonic systems. By refining our understanding of these mechanisms, we can advance the development of more efficient and high-performance optoelectronic devices.

\section*{Acknowledgments}

This study was supported by a National Research Foundation (NRF) grant funded by the Korean Government (Nos. NRF-2024-00411892) and 2025 Hongik Research Fund.

\bibliographystyle{unsrt}

%\begin{thebibliography}{1}

%\end{thebibliography}
\bibliography{main}
%\bibliography{}
\newpage

\vspace{11pt}


%\begin{IEEEbiography}[{\includegraphics[width=1in,height=1.25in,clip,keepaspectratio]{hkim.png}}]{Hyuntai Kim} received the Ph.D. degree from the School of Electrical Engineering, Seoul National University (SNU), South Korea, in 2016. He subsequently held a Postdoctoral Fellowship at SNU, for one year. He joined the Optoelectronics Research Centre (ORC), University of Southampton, as a Research Fellow, in 2017. Since 2019, he has been with the Department of Electronic and Electrical Converged Engineering, Hongik University, as an associate professor. He is primarily interested in fiber optics, nanophotonics, and artificial intelligence. 
%\end{IEEEbiography}

%\begin{IEEEbiography}[{\includegraphics[width=1in,height=1.25in,clip,keepaspectratio]{syc.jpg}}]{Seong-Yong Cho} received the Ph.D. degree from the Department of Materials Science and Engineering, Seoul National University. He held a Postdoctoral fellowship at the University of Illinois at Urbana-Champaign with colloidal QD based optoelectronic applications. He joined as a faculty in the Department of Materials Science and Engineering at Myongji University. He recently moved to Hanyang University in 2023 as an associate professor. His research interests are optoelectronic applications of nanomaterials and characterization of thin films.

%\end{IEEEbiography}

\vfill

\end{document}